\documentstyle[prb,aps,epsfig,twocolumn]{revtex}

\title { Magnetic fluctuations in frustrated Laves hydrides R(Mn$_{1-x}$Al$_{x}$)$_{2}$H$_{y}$} 

\author{P. Cadavez-Peres$^1$, I. Mirebeau$^1$, R. Kahn$^1$, I.N. Goncharenko$^{1,3}$, E. Vincent$^2$, and O.L. Makarova$^3$}

\address{1- Laboratoire L\'eon Brillouin, CEA-CNRS, CE Saclay, 91191 Gif sur Yvette, France}
\address{2- Service de Physique de l'Etat Condens\'e, CE Saclay, Orme des Merisiers, 
91191 Gif sur Yvette, France}
\address{3- Russian Research Center `Kurchatov Institute', 123182 
Moscow, Russia}

\date{\today, submitted to Phys. Rev. B.}

\begin{document}
\twocolumn[\hsize\textwidth\columnwidth\hsize\csname @twocolumnfalse\endcsname
\maketitle
\begin{abstract}
  By neutron scattering, we have studied the spin correlations and spin fluctuations in frustrated Laves hydrides, where magnetic disorder sets in the topologically frustrated Mn lattice. Below the transition towards short range magnetic order, static spin clusters coexist with fluctuating and alsmost uncorrelated spins. The magnetic response shows a complexe lineshape, connected with the presence of the magnetic inhomogeneities. Its analysis shows the existence of two  different processes, relaxation and local excitations, for the spin fluctuations below the transition. 

\end{abstract}
]
\section{Introduction}                                                                                                                              

There is now a growing interest in studying magnetic fluctuations in
the paramagnetic phase of strongly correlated electron systems.  An
original behavior in the dynamical susceptibility, where the
magnetization density does not follow the single exponential decay
predicted by Landau-Fermi theory is often observed.  In itinerant
magnets, non Fermi-liquid behavior is often associated with temperature anomalies of
the thermodynamical properties, like specific heat and resistivity. In Heavy-Fermions and
superconductors, it could be the precursor a quantum phase transition around zero temperature\cite{Sachdev99}.  

In paramagnets with localized spins, a non Fermi liquid behavior of
the dynamical susceptibility could be also expected due to strong
magnetic disorder, since magnetic inhomogeneities associated with
disorder and frustration yield a distribution of relaxation times. 
However, in usual spin glasses with localized spins, although such
distribution indeed exists, the excitation spectrum in the
paramagnetic phase usually remains much narrower than in itinerant
magnets, yielding typical energies of the fluctuating spins below 1
meV.

The intermetallic Laves compounds RMn$_2$ have two main
characteristics \cite{Ballou87,Shiga88a,Ritter91}: i) they are at the
frontier between localized and itinerant magnetism, the instability of
the Mn spins being governed by the distance between first neighbor Mn
pairs, and ii) they show a topological frustration of the pyrochlore
Mn lattice for antiferromagnetic first
neighbor Mn-Mn interactions. When R is a magnetic rare earth, the diamond-like R lattice interacts with the frustrated Mn lattice. Varying Mn-Mn interatomic distance d allows one to encompass the critical distance d$_{0}$ between localized and itinerant Mn magnetism. For d $>$ d$_{0}$, spontaneous Mn moments are present and the frustrated Mn-Mn interactions dominate, whereas for d $<$ d$_{0}$ Mn moments are induced by rare earth moments and the topological frustration is suppressed. Changing interatomic distances is realized by chemical substitution\cite{Ballou87,Shiga88a,Ritter91} or by applying pressure \cite{Mirebeau01}. 

 Introducing hydrogen in interstitial
sites also modifies the magnetic properties \cite{Vajda95}.  In
Laves compounds, hydrogen expands the lattice, favoring the
localization of the Mn moments, but also modifies the frustration by
making Mn-Mn interactions nonequivalent.  Therefore, ordered H atoms
release the topological frustration, yielding long range ordered
magnetic structures
\cite{Goncharenko97,Goncharenko99,Figiel98,Latroche00}.  On the other
hand, H disorder induces a random variation of the exchange interactions and therefore magnetic disorder\cite{Mirebeau00,Cadavez02}.

In Y(Mn$_{1-x}$Al$_{x}$)$_{2}$H$_{y}$ and
Y(Mn$_{1-x}$Fe$_{x}$)$_{2}$H$_{y}$, we studied such phases by magnetic
neutron diffraction\cite{Mirebeau00,Cadavez02}. Surprisingly, and in
contrast with usual spin glasses, we found that these disordered magnetic phases
combine short spin correlation lengths (around 10-20 {\AA}) with
 high transition temperatures (100-250 K), in the same range as in the ordered parent compound.  The
behavior of the short range order parameter was intermediate
between that expected for a canonical Edwards-Anderson spin glass, and
the Brillouin curve for a non frustrated magnet.

Up to now the study of the spin fluctuations and magnetic transition
in such topologically disordered phases has received little attention. 
A previous study of Y(Mn$_{0.9}$Al$_{0.1}$)$_2$ single crystal by inelastic neutron
scattering \cite{Motoya91}, where interatomic Mn distances are close
to the instability threshold, revealed very interesting features with
regards to canonical spin glasses. Namely, it pointed out the
existence of a large energy range of the magnetic fluctuations (up to
about 40 meV), well above the typical fluctuation energy in spin
glasses, and close to that observed in other itinerant magnets
\cite{itin:betaMn,itin:YSc}. This large energy range was attributed to the
influence of both topological frustration and itinerant character of
the Mn moments, without any way to clear up their respective
roles.

Studying  the magnetic excitations yields precious information about the microscopic nature of the magnetic ground state and ordering transition. Especially, it could help to explain why the transition temperature remains high in spite of the strong disorder. Inelastic neutron scattering is the most straightforward way to probe such excitations. We have studied the spin fluctuations in
 R(Mn$_{1-x}$Al$_{x}$)H$_{y}$ compounds
(R = Y or Tb) by time of flight neutron scattering, together the steady-state correlations by neutron
diffraction, and the low field susceptibility by static magnetization
measurements. In all compounds, H disorder was achieved by
substituting a small amount of Al on the Mn sites.  By choosing
appropriate compounds, we checked the influence of magnetic
moments on the R sites, hydrogen concentration and hydrogen disorder. The results suggest an original picture of a `cluster glass' magnetic state.

\section{Sample description}

 Powdered samples of R(Mn$_{1-x}$Al$_x$)$_2$D$_y$, with (R = Y,
Tb), Al content $x$=0.09, and different deuterium contents 
were prepared using the technique described in Ref. \onlinecite{Mirebeau00},
and characterized by X-ray diffraction.  All samples showed a single
chemical phase, besides a very small amount of RH$_{3}$ impurity phase in
some cases.  
We mostly studied three
disordered samples, Tb(Mn$_{0.9}$Al$_{0.1}$)$_2$D$_y$, ($y$=1 and
$y$=4), and Y(Mn$_{0.9}$Al$_{0.1}$)$_2$D$_y$ ($y$=4).  
Exact D contents are given in Table \ref{table1}. In most cases, the deuterium
isotope was chosen instead of hydrogen to decrease the incoherent
background. The small Al content induces hydrogen
disorder and breaks down the long range magnetic order
\cite{Mirebeau00}. The Tb ion with high magnetic moment yields a
strong enhancement of the magnetic fluctuations (by about a factor
10), yielding an excellent precision on the lineshape of the dynamical
susceptibility. The influence of this second magnetic ion on the spin
correlations and spin dynamics was checked in comparison with
Y(Mn$_{0.9}$Al$_{0.1}$)$_2$D$_4$ where Y is non magnetic. The amount of hydrogen (or deuterium) introduced ($y$=1 or $y$=4) allows to vary interatomic distances in a large range.  In this range, the d$_{\rm Mn-Mn}$ distance (2.77 {\AA}
for Tb(Mn$_{0.9}$Al$_{0.1}$)$_2$D$_1$, and 2.87 {\AA} 
for Tb(Mn$_{0.9}$Al$_{0.1}$)$_2$D$_4$) remains above the magnetic
instability threshold for the Mn moments (d$_{0}$$\simeq$2.66 {\AA}).
 
In comparison with the disordered compounds we studied the spin
fluctuations in ordered TbMn$_2$D$_4$, TbMn$_2$H$_4$, and YMn$_2$H$_4$. Their simple long range
antiferromagnetic structure, which is determined by the H
superstructure \cite{Goncharenko97,Goncharenko99}, is
a reference to check the influence of hydrogen or deuterium disorder on the
spin fluctuations.
 
\section {Neutron diffractiion measurements}

We studied the magnetic correlations on the powder diffractometer G6-1
of the reactor Orph\'ee (with an incident neutron wavelength of 4.734
{\AA}). Temperature varied between 350 K and 10 K. A
spectrum measured in the paramagnetic range was subtracted to extract
the magnetic correlations. The magnetic spectra of the Al substituted
samples show diffuse
magnetic peaks, corresponding to short range antiferromagnetic correlations (Fig. 1).  The peak positions are indexed with the same propagation vector as in the
ordered parent compound RMn$_2$D$_4$. Using
the Fullprof program \cite{Rodriguez93}, we refined the data
assuming short range ordered antiferromagnetic regions of diameter L$_c$ with ordered moments $\mu_{\rm R}$ and $\mu_{\rm Mn}$, keeping the same type of order as in Ref.\onlinecite{Goncharenko99}. Results
 are given in Table \ref{table1}. 
 We find correlated regions of extension  L$_c$ between 15 and 30 \AA.  L$_c$ is
temperature independent, but varies with H content and R substitution. 
The presence of magnetic Tb ions increases  L$_c$ with respect to
non magnetic Y. It also induces {\it ferromagnetic} correlations on a
shorter length scale (about 6 {\AA}). These ferromagnetic correlations, shown by an increase of the magnetic scattering at low angles coexist with the antiferromagnetic ones. The short range ordered moments
remain well below the moments in the long range ordered
samples ($\mu_{\rm Mn}$=4 $\mu_B$ and $\mu_{\rm Tb}$=9 $\mu_B$ in
ordered TbMn$_2$D$_4$ for instance). In spite of the strong magnetic disorder, the transition
temperatures T$_{\rm SRO}$ towards the short range ordered state are in the range 100-250 K, therefore not
much decreased with respect to the  N\'eel temperature of the long range ordered state
(T$_{\rm N}$ =300 K in TbMn$_2$D$_4$ and 330 K in YMn$_2$D$_4$). This was also the case of other disordered hydrides
previously studied \cite{Cadavez02}. A detailed description of the
short range magnetic order in the
R(Mn$_{0.9}$Al$_{0.1}$)$_{2}$H$_{y}$ family will be given later.

\section  {magnetic measurements}

The magnetization was measured with a SQUID magnetometer, under static
magnetic field of 10 Oe, in the zero field cooling (zfc) and field
cooling (fc) conditions. Typical curves are shown in Fig. 2. 
In all disordered samples, magnetic irreversibilities start around 230-280 K, namely 
above T$_{\rm SRO}$. In the Y-samples, where only Mn sublattice is magnetic, the fc curve continuously increases with decreasing temperature, well above the zfc curve, which shows a broad maximum 
 then flattens or decreases. The magnetization of the two Y-samples have the same order of magnitude in the paramagnetic regime, but in the irreversibility region, the magnetization is much smaller for the sample with low hydrogen content. The temperature dependence of both fc and zfc curves is reminiscent of the behavior in cluster glasses \cite{Koyano94,Pejakovic00,GomezSal02},
where magnetic order occurs within spatially limited regions, and the relative orientations 
of the superparamagnetic-like spin clusters depend on the cluster interaction field, anisotropy 
field, and thermal activation \cite{Nozar87}. In the Tb-samples with two 
magnetic sublattices, this cluster glass behavior is superimposed 
on a global paramagnetic component, with a variation close to 1/T.  
This second component should be related with the strong thermal 
fluctuations of the Tb moments, which become progressively polarized by the 
Tb-Mn exchange field as temperature decreases \cite{Goncharenko99}. 
In all the short range ordered samples, anomalies are observed in the low temperature
range (30-50 K), as shown by a small maximum in the zfc curve, and an
upturn of the fc curve. Similar anomalies were predicted and observed in some
cluster glasses, when the local anisotropy fields start to be
significant \cite{Nozar87}.
They suggest the presence of small superparamagnetic clusters which could fluctuate 
down to low temperatures, having their anisotropy fields perpendicular to the interaction 
fields\cite{Nozar87}. In Y(Mn$_{0.9}$Al$_{0.1}$)$_{2}$D$_{4}$, the
low temperature anomalies result in a plateau on
the fc curve. This plateau, whose origin is unexplained, reaches only about 3 $\%$ of the maximum expected due to demagnetization effect.

In comparison with the short range ordered compounds, 
the magnetization of YMn$_2$D$_4$ and TbMn$_2$D$_4$ with long range antiferromagnetic order is shown in the insets.  
In both samples, the magnetic irreversibilities start abruptly at the N\'eel temperature 
T$_{\rm N}$ measured by neutrons \cite{Goncharenko97,Goncharenko99}. In YMn$_2$D$_4$, 
a magnetization plateau is observed below T$_{\rm N}$, with a 
value of about 3 $\%$ of the demagnetization value. This plateau, 
rather unusual for an antiferromagnet, is presumably connected with the
first order character of the transition, yielding a spin blocking 
induced by hydrogen order and ordered antiferromagnetic moment 
which remains close to saturation almost up to T$_{\rm N}$. 
TbMn$_2$D$_4$ shows a similar behavior, 
with an additional paramagnetic component. In both samples, 
the low temperature anomalies observed in the short range 
ordered samples are absent. 
   
\section{Inelastic neutron scattering}

\subsection{experimental set up and data analysis}

The experiments were performed in the time of flight spectrometer
Mibemol of the reactor Orph\'ee, with an incident wavelength of 5
{\AA} (incident energy  of 3.27 meV). The energy resolution (FWHM=0.15 meV) and the
efficiency of the detectors were determined by measuring a vanadium
standard of the same plate-like shape as the samples.  After
background and absorption corrections, the neutron cross sections were
calibrated in absolute scale using the vanadium spectrum. 
Measurements were performed {\em versus} temperature, using either a closed
cycle nitrogen flow cryo-oven (between 350 K and 100 K) or a helium
flow cryostat (between 300 K and 20 K). The maximum temperature of
350 K was chosen to avoid any decomposition of the samples due to
hydrogen migration. The temperature range allowed us to study the
spin fluctuations in the paramagnetic range, and to follow their
evolution below the transition. The range of the elastic scattering 
vector is the same as in the diffraction measurements.
 
The scattering cross section for magnetic quasielastic neutron
scattering can be expressed in a general form 

\begin{equation}
S(q,\omega)=\frac{1}{\pi}n(\omega){\chi}_{q}''(\omega)    
\label{eq:TRL}
\end{equation}
where $n(\omega)$ is the Bose factor taking into account the thermal population factor of the fluctuations, and ${\chi}_{q}''(\omega)$ the
dynamical susceptibility yielding the response of the system.  Within
linear response theory, ${\chi}_{q}''(\omega)$ is the product of
the static transverse susceptibility ${\chi}$(q,$\omega$=0), and the
spectral function $F(\omega)$. In the simplest case of a
single exponential decay (SED), there is a single energy $\Gamma$ (or
equivalently a single relaxation rate $\tau$=1/2$\pi\Gamma$, taking
$\hbar$=1) for the magnetic fluctuations, so that the spectral
function is expressed as $F(\omega)=\frac{1}{\pi}\frac{\Gamma}
{\Gamma^{2}+\omega^{2}}$. In classical spin glasses like CuMn, the spectral response was expressed by a distribution of Lorentzians correspondig to a distribution of relaxation rates \cite{Murani78,Murani85}. Still, the quasielatic scattering cross section in the experimental energy range
was well fitted by a single Lorentzian lineshape\cite{Murani78}.
   
 This simple model could not account for our data, yielding strong misfits even in the paramagnetic range.  Therefore we considered explicitely an energy distribution in the spectral lineshape. We took two limiting forms for this distribution, namely a bimodal
distribution (model 1) and a uniform distribution (model 2).  Model 1
assumes two typical energies $\Gamma_{1}$ and $\Gamma_{2}$
($\Gamma_{1}$$<$$\Gamma_{2}$) for the spin fluctuations.  It was used
previously to analyze the spin dynamics in cluster glasses
\cite{Scheuer77,Mirebeau96}.  Model 2 assumes that the energies of the spin
fluctuations are distributed with an equal probability between lower
and higher energy bounds $\Gamma_{1}$ and $\Gamma_{2}$. It was  recently proposed by Bernhoeft \cite{Bernhoeft01} within a general model of
the dynamical phase inhomogeneities in non Fermi liquid systems.

The quasielastic neutron cross section is written

- within model 1:

\begin{equation}
S(q,\omega)=\frac{\omega}{e^{\omega/kT}-1}\left(F_{1}(\omega)\chi_1(q) +
F_{2}(\omega)\chi_2(q)\right)
\label{eq:S(q,w)}
\end{equation}
where $\chi_1$ and $\chi_2$ are the static susceptibilities of the two
quasielastic contributions and the spectral functions are respectively

\begin{equation}
F_{1}(\omega)=\frac{1}{\pi}\frac{\Gamma_{1}}
{\Gamma_{1}^{2}+\omega^{2}} \;\;\;\;
F_{2}(\omega)=\frac{1}{\pi}\frac{\Gamma_{2}}
{\Gamma_{2}^{2}+\omega^{2}}
\label{eq:F(w)}
\end{equation}
with $\Gamma_{1}$, $\Gamma_{2}$ being the corresponding energy linewidths.
The static susceptibility is defined as $\chi$=$\chi_1$+$\chi_2$.

- within model 2:

\begin{equation}
S(q,\omega)=\frac{\omega}{e^{\omega/kT}-1} \, F(\omega) \, \chi(q)
\label{eq:S(q,w)new}
\end{equation}
where $\chi$ is the static susceptibility and $F(\omega)$ the spectral function 
written as 

\begin{equation}
F(\omega)=\frac{1}{\pi} \frac{1}{\ln(\Gamma_{2})-\ln(\Gamma_{1})} \,
\frac{1}{\omega} \left(\arctan(\frac{\omega}{\Gamma_{1}}) -
\arctan(\frac{\omega}{\Gamma_{2}})\right)
\label{eq:F(w)new}
\end{equation}
where $\Gamma_{1}$ and $\Gamma_{2}$ are respectively the low and high energy limits
of the energy distribution.

In the above expressions, $\omega$$>$0 corresponds to a neutron energy
gain.  The prefactors in equations (\ref{eq:F(w)}) and 
 (\ref{eq:F(w)new}) 
 normalize to
unity the energy integration of the spectral function. Therefore, in
the high temperature limit $\omega$$\ll$kT (valid here in the
paramagnetic range and down to about 50 K), the quantity T$\chi$ is
equal to the energy-integrated intensity of the quasielastic signal.
 The fitted energy widths do not depend on the normalisation factor. 

In the limit of a Curie-Weiss behavior, the neutron susceptibility is
related to the effective fluctuating moment by the expression:
$\chi$(q=0,T)=$A\frac{m_{eff}^2 \mu_{B}^2}{3kT}$. $A$ corresponds to
the calibration of the neutron cross section in absolute scale.  This
expression provides an effective moment value, to be compared with the
value from bulk magnetic measurement.

The total cross section is written as

\begin{equation}
S^{\rm total}(q,\omega) = c_{el} \delta(\omega) + S(q,\omega)
\label{eq:}
\end{equation}

The  neutron intensity 
$I$ detected {\em versus} time of flight $t$ at the scattering angle $2\theta$ is proportional to the quantity:

\begin{equation}
 I(\theta,t) \propto \frac{1}{t^{4}} S^{\rm total}(q,\omega) 
\label{eq:I}
\end{equation}

The neutron intensity was convoluted with the experimental resolution and fitted to each
individual time of flight (TOF) spectrum, with $c_{el}$ (intensity of the elastic
peak), $\Gamma_{1}$, $\Gamma_{2}$, and $\chi_{1}$, $\chi_{2}$ (model
1) or $\chi$ (model 2) as parameters.  A very small linear background
was also added in the fit.  The fit was performed in the energy range (-20, 2) meV, where the phonon contribution is 
small and could be neglected.


\subsection  {results}

{\bf a) short range ordered samples}

We first show the results in Tb(Mn$_{0.9}$Al$_{0.1}$)$_2$D$_y$ ($y$=1
and $y$=4).  In spite of their different D contents, ordering
temperatures and magnetic correlations lengths, the behavior of the
magnetic fluctuations is the same for the two samples. 
Typical TOF spectra are shown in Figs. 3 and 4 for temperatures respectively above and below T$_{\rm SRO}$. Note that due to the $t^{-4}$ factor of equation (\ref{eq:I}), the maxima of the quasielastic signals are shifted with respect to the elastic one. In spite of this distortion, the time dependent plot were preferred to the energy dependent ones (in inset), since they provide a better estimation of the fit quality at low energies.

 Whatever the model chosen for the spectral function, the general evolution with temperature can be described as follows.
 Above T$_{\rm SRO}$, the susceptibility follows a Curie law, and the
quasielastic integrated intensity T$\chi$ is temperature independent
(Fig.  5).  Below T$_{\rm SRO}$, we observe a gradual transfer of
intensity from the quasielastic signal to the elastic one, the sum
of the two contributions being conserved with temperature. 
The high energy signal of width $\Gamma_{2}$ dominates, 
with a weight of 80$\%$ of the suceptibility at high temperature. The energy
linewidths have a rather original behavior (Fig. 6). 
 Above T$_{\rm SRO}$, $\Gamma_{1}$ and $\Gamma_{2}$ are temperature independent.  Below
T$_{\rm SRO}$, the small linewidth $\Gamma_{1}$ starts to decrease
whereas the large linewidth $\Gamma_{2}$ {\it increases} with
decreasing temperature.  This behavior is the same for all q values.
 In the paramagnetic regime, the quasielastic cross section is well fitted by
the two models.  Below T$_{\rm SRO}$, model 1 still yields very good fit
of the data, whereas model 2 starts to show some misfits.  They occur
mainly below 80 K where the linewidths $\Gamma_{1}$ and $\Gamma_{2}$
are very different.  We have therefore restricted the use of model 2
to temperatures above 80 K, where this model was meaningful. 
In the paramagnetic range, the two models give an energy linewidth in 
the same energy range, with $\Gamma_{1}$ around 0.1 meV, and 
$\Gamma_{2}$ around 3-5 meV. 

The q dependence of the susceptibility and quasielastic widths is
shown in Fig. 7.  The susceptibility $\chi$ shows a weak and broad modulation with respect to the form factor of the free ion. This modulation remains the same below T$_{\rm
SRO}$. The magnetic correlations observed below T$_{\rm
SRO}$ by neutron diffraction correspond to a modulation of the  {\it elastic} component. 
The energy linewidths are almost q-independent, both above and
below T$_{\rm SRO}$.  The 
Tb(Mn$_{0.9}$Al$_{0.1}$)$_2$D$_1$ sample, where the magnetic correlations are enhanced, shows the same behavior.

In Y(Mn$_{0.9}$Al$_{0.1}$)$_2$D$_{4}$, the magnetic signal
coming from Mn ions only is much smaller, 
 but still shows the complex lineshape described above.  
The main difference with the Tb samples concerns the temperature
dependence of the large energy linewidth $\Gamma_{2}$, which now strongly {\it
decreases} below T$_{\rm SRO}$ (Fig.  8). This is observed whatever the
model used to describe the spectral function.
 
\vspace{2mm}

{\bf b) long range ordered samples}

In comparison, we show the magnetic fluctuations in
TbMn$_2$D$_4$, TbMn$_2$H$_4$, and YMn$_2$H$_4$ where long range
magnetic order occurs at T$_{\rm N}$ through a first order
magnetostructural transition. 
Above T$_{\rm N}$, the magnetic fluctuations are very similar in the ordered and in the disordered samples. The main effect of the LRO is a strong suppression of the quasielastic intensity below T$_{\rm N}$. For example, at T= T$_{\rm N}$/2 the intensity of the magnetic fluctuations is decreased by a factor 3 with respect to its value in the paramagnetic state in the ordered Tb samples, and by a factor 5 in ordered  YMn$_2$H$_4$, whereas it decreases by 30$\%$ in disordered Tb(Mn$_{0.9}$Al$_{0.1}$)$_2$D$_1$ and only 8$\%$ in Y(Mn$_{0.9}$Al$_{0.1}$)$_2$D$_{4}$. The second important difference is the occurence of a well defined localized excitation in the ordered Tb samples below T$_{\rm N}$.

In Fig.  9, we show two typical TOF spectra for TbMn$_2$D$_4$ at 340 K and 160 K, namely above and below
T$_{\rm N}$. At 340 K, the spectral lineshape is well fitted by a
bimodal distribution (model 1), with typical energy widths of 0.5 and
2 meV, therefore comparable to the energy widths of the disordered
samples at the same temperature. Below T$_{\rm N}$, a purely
inelastic signal appears in the TOF spectra, coexisting with the
quasielastic one. This new component was fitted by a Gaussian lineshape, 
of energy E$_{0}$ and width $\Gamma_{\rm inel}$. Its intensity (fig. 10a) shows a broad maximum versus temperature. The energy E$_{0}$ slightly
increases from 6 to 7 meV when temperature decreases (Fig 10b), with constant
energy width $\Gamma_{\rm inel}$ of 2 meV.
The quasielastic signal strongly decreases just below T$_{\rm N}$, but still persist in the ordered state. The large energy width $\Gamma_{2}$ increases with decreasing T, whereas the small energy width $\Gamma_{1}$ is about temperature independent. The sum of the three contributions (elastic, quasielastic and inelastic) is conserved with
temperature.

In TbMn$_2$H$_4$, the inelastic and quasielastic signals are the
same as in the TbMn$_2$D$_4$. 
The elastic component is enhanced by about a factor 10 in TbMn$_2$H$_4$, due
to the large incoherent contribution of the hydrogen isotope.  As
expected from previous inelastic neutron studies of Laves hydrides
\cite{Campbell98}, the dynamics of hydrogen ordering cannot be
resolved in the present experiment. In other words, hydrogen (or deuterium) is frozen at the time scale of the neutron experiment (t=9$\times$$10^{-12}$s), and
there is no interference between H or D diffusion and the magnetic
fluctuations.
 
Finally, in ordered YMn$_2$H$_4$ (Fig 11), the quasielastic signal above  T$_{\rm N}$ also consists of two components. Below T$_{\rm N}$, it becomes very small but can still be detected down to 100 K. The energy width in the paramagnetic state are about the same as in the disordered Y(Mn$_{0.9}$Al$_{0.1}$)$_2$D$_{4}$, but here they do not vary much with temperature. The inelastic localized mode observed in the ordered Tb samples was not observed. 

\section{Discussion}

\subsection{paramagnetic regime}

In the high temperature range where the samples show a Curie-Weiss behavior, the effective moments deduced from the integration of the
quasielastic neutron cross section agree with the value deduced from
magnetization (table \ref{table2}).  This means that most of the spin
fluctuations are probed by the experimental time window, and rules out
the presence of high energy spin fluctuations (in the 50 meV range or
above), similar to that observed in YMn$_2$ \cite{YMn2fluct1,YMn2fluct2},
Y(Mn$_{0.9}$Al$_{0.1}$)$_2$ \cite{Motoya91} or other itinerant
systems.  Therefore, all samples are in the localized moment limit,
even at low H(D) content, and the wide energy distribution seen
in both Tb and Y samples should be connected with peculiar features of
the magnetic disorder.  At high temperature, for localized spins
coupled by near neighbor interactions, the energy linewidth of the
magnetic fluctuations is related to the values of the exchange
constants \cite{Marshall}. This is the case of the present samples,
where the anisotropy energy is much smaller than the exchange energy
\cite{Cadavez00}. Within molecular approximation, the lineshape of
the response function is expected to be intermediate between Gaussian
and Lorentzian, and its second moment, which measures the quasielastic
width is given for a polycrystalline sample by

\begin{equation}
\overline{\omega^2}=
(8/3)S(S+1)\sum_{i}J^2({\rm R}_{i})(1-\sin({\rm q}{\rm R}_{i})/({\rm q}{\rm R}_{i}))
\label{eq:largeur}
\end{equation}

In this model, the energy linewidth $\sqrt{\overline{\omega^2}}$ is
temperature independent, in agreement with experiment, and in contrast
with the behavior of itinerant systems where the linewidth increases
linearly with temperature (Korringa behavior).  Assuming first
neighbor interactions only (R$_{1}$=2.86 {\AA}), the smooth variation of
the linewidth predicted by equation (\ref{eq:largeur}) above 1 {\AA}$^{-1}$
is also consistent with experiment (Fig.  7b),
although the expected decrease at low q values is not seen.

The distribution of linewidths may be connected with a distribution of
exchange constants, induced by the frozen hydrogen or deuterium disorder.  As
discussed previously for the ordered compounds YMn$_{2}$D$_{4}$, the
presence of a H(D) atom near a Mn-Mn pair modifies the amplitude, and
even the sign, of the first neighbor exchange interaction
\cite{Goncharenko99}.  The various environments of the Mn atoms having
zero or one H(D) atom nearby the Mn pairs should be at the origin of the
linewidth distribution.  In the Tb samples, the magnetic fluctuations
are dominated by the Tb moments of much larger magnitude.  The
distribution of linewidths therefore mostly measures the distribution
of Tb-Mn interactions, showing that they are also sensitive to the
number of H(D) near neighbors atoms.  This is also consistent with the analysis of the
ordered structure\cite{Goncharenko99}. 
Taking $S_{\rm Mn}$=2, and summing equation (8) over the 6 first
neighbors of a Mn ion, we obtain from the two energy widths in Y(Mn$_{0.9}$Al$_{0.1}$)$_2$D$_{4}$, low and high limits for the Mn-Mn
exchange constant $\sqrt{\overline{J^2}_{\rm
Mn-Mn}}$ equal to 0.03 and 0.3 meV. A similar evaluation for
Tb(Mn$_{0.9}$Al$_{0.1}$)$_2$D$_{4}$ ($S_{\rm Tb}$=3) yields limit values for the the Tb-Mn
exchange constant $\sqrt{\overline{J^2}_{\rm Tb-Mn}}$ equal to 0.03 and
 0.1 meV.
 
\subsection{ordering transition}

In the long range ordered compounds, the transition is of
first order, whereas in the short range ordered compounds, it is close to second order\cite{Goncharenko97,Mirebeau00}. As shown previously, H(D) order is the key parameter which controls the frustration. In the LRO samples, the first order transition is directly connected with the formation of H(D)
superstructure which relieves the frustration of the Mn lattice. In contrast, in the disordered samples, the second order transition occurs in a range of temperature where the slowing down the H(D) diffusion yields a more progressive change from thermal to frozen disorder. In both cases, the
present results show that the transition is characterized by a deep
change in the magnetic fluctuations, namely, i) a transfer of intensity
from the quasielastic to elastic signal connected with the onset of {\it static}
correlations, and ii) a change in the behavior of the quasielastic widths,
which start to vary with temperature. 


The transition temperature T$_{\rm SRO}$ remains in the range 100-250 K, therefore is not much decreased with respect to  the N\'eel temperature of the ordered compounds (300-330 K). Such high transition temperature is often observed in cluster glasses \cite{Itoh94}.  It
contrasts with classical spin glasses where the spin glass temperature is typicallty ten times smaller than the transition temperature of the ordered
parent compound\cite{Itoh94,Shiga85,Menshikov94}. The average
exchange constant $J_{\rm Mn-Mn}$ deduced from T$_{\rm N}$ in the mean
field approximation (T$_{\rm N}$=$\sum_{i}$$\frac{2}{3} J_{\rm Mn-Mn} S_{\rm Mn}(S_{\rm Mn}+1)$, summing over the 6 first neighbors of a 
Mn ion) yields $J_{\rm Mn-Mn}$=0.9 meV, which is about 3 times
higher than the upper limit for $\sqrt{\overline{J^2}_{\rm Mn-Mn}}$
deduced from the linewidths in the paramagnetic regime.  
Obviously, such evaluations only provide an order of magnitude of the exchange constant since they neglect the influence of inhomogeneity and frustation. 

A simple picture in terms of a cluster glass can account for the observed decrease of T$_{\rm N}$ with respect to the ordered state. One assumes clusters of diameter L$_c$ (where L$_c$ is the correlation length) consisting of an ordered core surrounded by a disordered layer of thickness d (where d=2.8{\AA} is the first neighbor distance). In first approximation, the decrease of T$_{\rm N}$ is given by the percentage disordered spins, assuming that they do not contribute to the average exchange interaction.  For the Tb samples (T$_{\rm N}$=300K in ordered TbMn$_2$D$_4$), this simple picture yields short range ordering temperatures of 112K (y=4) and 161K (y=1), in relatively good agreement with the experimental values (135K and 200K respectively).  A better evaluation would require to define different 
inter and intracluster exchange interactions, as proposed in mean 
fields models of cluster glasses \cite{Soukoulis78}. 
  
The coexistence
of highly frustrated and low frustrated spins yields a large
distribution of energy barriers, as observed in cluster glasses or
inhomogeneous superparamagnets.  The magnetization data support this
qualitative picture, showing that some spins could freeze either above
T$_{\rm SRO}$ (up to the N\'eel temperature of the ordered 
compound), or well below T$_{\rm SRO}$ down to the low T range.
 
\subsection{spin dynamics below the transition}

As temperature decreases, more and more spins freeze and contribute to
the static clusters. The fluctuating spins remain almost uncorrelated like in the paramagnetic regime.  In
Y(Mn$_{0.9}$Al$_{0.1}$)$_2$D$_{4}$, the decrease of the two
quasielastic widths suggests that these uncorrelated spins relax
between several metastable configurations, with typical times
($\tau$=1/2$\pi\Gamma$) increasing with decreasing temperature.

In the Tb samples, the formation of the static clusters is associated with the emergence of two types of spin dynamics : a relaxation process involving weakly coupled and almost uncorrelated spins, as shown by the decrease of the small quasielastic width $\Gamma_{1}$, and an excitation process as shown by the
increase of the large quasielastic width $\Gamma_{2}$. Here the magnetic signal is
dominated by the Tb spins. In ordered TbMn$_{2}$D$_{4}$, a local mode of energy E$_{0}$ is clearly observed below T$_{\rm N}$, and is attributed to local fluctuations of the Tb spins in the static Mn molecular field. Such mode is predicted and observed in ordered RFe$_{2}$ or RAl$_{2}$ Laves
phases \cite{Paolasini,Rhyne}. Its energy is expected to increase smoothly with temperature like the ordered Mn moment(E$_{0}$ $\propto$ $J_{\rm Tb-Mn}$ $S_{\rm Mn}$). In the short range ordered Tb compounds, one should also expect a similar inelastic mode, but broadened by the distribution of exchange interactions and situated at a lower energy, due to the lower value of the short range ordered Mn moment. Therefore, this mode could be hidden by the large quasielastic signal. In order to check this assumption, we have performed for Tb(Mn$_{0.9}$Al$_{0.1}$)$_2$D$_4$ the same fit as for ordered TbMn$_{2}$D$_{4}$. It yields a slight improvement of the fit quality, with an inelastic contribution of about
 10$\%$ of the quasielastic one. The energy E$_{0}$ is smaller than in TbMn$_{2}$D$_{4}$ and increases more rapidly with decreasing temperature (from 3 to 5 meV), as expected from a simple model. The quasielastic widths, especially the low energy width $\Gamma_{1}$ are not much affected.

 This analysis points out the strong similarities between the spin fluctuations in the short range and long range ordered Tb compounds. It also suggests that the large quasielastic signal and the inelastic one have the same nature, and we note that their characteristic energies E$_{0}$ and $\Gamma_{2}$ vary in the same way with temperature. The onset of short range order weakens the energy of the inelastic mode, and changes the balance between the two contributions. The persistence of a quasielastic signal in the ordered state, which is not observed in RFe$_{2}$ Laves phases, is presumably related to the influence of hydrogen or deuterium. The lowering of symmetry due to the H(D) superstructure\cite{Goncharenko99} might create lower energy levels which cannot be distinguished from the quasielastic scattering. The residual disorder, since one H(D) site is halfly occupied, could also play a role. Whatever the exact mechanism, we have observed local excitations of low energy which develop as soon as a static magnetic order freezes in, and are almost insensitive to the correlation length. A better analysis of these excitations would require higher resolution measurements on a single crystal.

\section {Conclusion}

In the Laves compounds R(Mn$_{0.9}$Al$_{0.1}$)$_2$D$_{y}$ with deuterium or hydrogen disorder, a transition
towards short range magnetic order is probed by neutron scattering. It is shown by the occurrence of static
clusters of a few tens of Angstroms, with a size independent of temperature. 
The transition temperature T$_{\rm SRO}$ is high in spite of the small cluster
size, and the transition is close to second order, in contrast with the first order transition in the long range ordered compounds. 
 In the
paramagnetic range, the complex lineshape of the spectral
function suggests a distribution of exchange constants, possibly
related to the onset of nonequivalent magnetic bonds and the influence
of frozen H(D) configurations. Below T$_{\rm SRO}$, several types
of fluctuations are observed, depending on the presence of the Tb magnetic ion. They are attributed either to the relaxation
of almost uncorrelated spins, or to local spin excitations within the frozen clusters. The spin fluctuations in the short range and long range ordered compounds show strong similarities.
The overall behavior and the macroscopic magnetic properties are interpreted in terms of a cluster glass, where short
range ordered regions with local H(D) order and low
frustration coexist with strongly frustrated and uncorrelated spins. 
It would be interesting to know if the observed spin fluctuations arise
from the peculiar coupling of hydrogen and magnetic lattices, or if they could
be observed in other cluster glasses or disordered magnets with topological frustration.

\acknowledgements
We thank N. Bernhoeft, M. Hennion, P.A. Alekseev and J.C. G\'omez Sal for many
interesting discussions.  P. C.-P. is supported by Funda\c{c}\~ao 
para a Ci\^encia e a Tecnologia, Portugal, through the research grant
PRAXIS\,XXI/BD/20334/99. This work was partly supported by the Russian foundation for Basic Research, Grant N. 02-02-06085,and the Russian State Program 'Neutron Investigations of Condensed Matter'.


\vspace{6mm}

{\large {\bf Figure captions}}

\vspace{3mm}

Fig.1: 
Magnetic neutron diffraction spectra at 10 K for
Tb(Mn$_{0.9}$Al$_{0.1}$)$_2$D$_y$ ($y$=4 and $y$=1) and
Y(Mn$_{0.9}$Al$_{0.1}$)$_2$D$_4$, measured on G6-1 diffractometer
($\lambda$ =4.734 {\AA}).  A spectrum measured in the paramagnetic
range was subtracted to extract the magnetic
contribution. For Tb(Mn$_{0.9}$Al$_{0.1}$)$_2$D$_4$, the first 
antiferromagnetic peak is not observed since magnetic moments are 
along direction $<$111$>$. Solid lines are fits of the data, yielding the short range ordered magnetic moments and correlations lengths given in Table 1.

\vspace{3mm}

Fig.2: 
Low field magnetization in Tb(Mn$_{0.9}$Al$_{0.1}$)D$_{y}$ ($y$=4 and
$y$=1), and in Y(Mn$_{0.9}$Al$_{0.1}$)D$_{y}$ ($y$=4 and $y$=1),
measured in a static field H=10 Oe, using zero field cooling (zfc) and
field cooling (fc) conditions. Insets: low field magnetization
of the long range ordered YMn$_2$D$_4$ and TbMn$_2$D$_4$ compounds. 
The transition temperature T$_{\rm SRO}$ towards short range order is shown by
arrows. Other arrows point out the low temperature anomalies.

 
\vspace{3mm}

Fig.3: 
Time of flight (TOF) spectrum of Tb(Mn$_{0.9}$Al$_{0.1}$)D$_{4}$ at
250 K (paramagnetic range), focusing on the quasielastic region.
Spectra were regrouped for detectors within an angular range 23.5--63.1
degrees.  Fits with model 1 (a) and model 2 (b) are shown by
solid lines. Inset: the corresponding neutron cross section {\em versus} energy transfer, and a schematic drawing of the distribution of energy widths for each model.

\vspace{3mm}

Fig.4: 
TOF spectrum of Tb(Mn$_{0.9}$Al$_{0.1}$)D$_{4}$ at
80 K (below T$_{\rm SRO}$), in an angular range 23.5--63.1
degrees. Fits with model 1 (a) and model 2 (b) are shown by
solid lines. Inset: the corresponding neutron cross section {\em versus} energy transfer and a schematic drawing of the distribution of energy widths for each model.
\vspace{3mm}

Fig.5:
Elastic, quasielastic intensity T$\chi$, and sum of the two
contributions {\em versus} temperature in Tb(Mn$_{0.9}$Al$_{0.1}$)$_2$D$_{4}$. 
TOF spectra have been regrouped within an angular range 23.5--63.1
degrees.  Models 1 and 2 give equivalent results.

\vspace{3mm}

Fig.6: 
Quasielastic widths $\Gamma_{1}$ and $\Gamma_{2}$ {\em versus}
temperature in Tb(Mn$_{0.9}$Al$_{0.1}$)$_2$D$_{4}$, obtained by fitting the data with model 1 (a) and model 2 (b).

\vspace{3mm}

Fig.7: 
Results obtained in Tb(Mn$_{0.9}$Al$_{0.1}$)$_2$D$_{4}$ with model 1.  a) Quasielastic
intensity {\em versus} q for two temperatures, 200 K and 80 K, above and
below T$_{\rm SRO}$ respectively.  Solid lines correspond to the squared
magnetic form factors of Tb and Mn ions. Dotted lines are guides to the eye.
 b) Quasielastic widths
$\Gamma_{1}$ and $\Gamma_{2}$ {\em versus} q at 200 K. Solid lines are
fits with the equation (\ref{eq:largeur}) described in text.  c)
Quasielastic widths {\em versus} q at 80 K.

\vspace{3mm}

Fig.8: 
Quasielastic widths $\Gamma_{1}$ and $\Gamma_{2}$ {\em versus}
temperature in Y(Mn$_{0.9}$Al$_{0.1}$)$_2$D$_{4}$, obtained with
model 1.

\vspace{3mm}

Fig.9: 
TbMn$_2$D$_4$ : TOF spectra at 340 K and 160 K, 
(T$_{\rm N}$ =300 K). In inset, the corresponding neutron cross section {\em versus} energy transfer. Solid lines are fits with model
1, adding an inelastic contribution below T$_{\rm N}$.

\vspace{3mm}

Fig.10: 
TbMn$_2$D$_4$ : a) variation of the integrated intensities {\em versus}
temperature, b) quasielastic widths $\Gamma_{1}$ and $\Gamma_{2}$ and
energy E$_{0}$ of the inelastic signal {\em versus} temperature.
\vspace{3mm}

Fig.11: 
YMn$_2$H$_4$ : a) variation of the integrated intensities {\em versus}
temperature, b) quasielastic widths $\Gamma_{1}$ and $\Gamma_{2}$ {\em versus} temperature.

\vspace{2cm}

{\large{\bf Table captions}}

\vspace{3mm}

\begin{table}[!b]
\begin{center}
\begin{tabular}{cccccc}
R & $y$ (exact value) &  T$_{\rm SRO}$ (K) &  L$_c$ 
(\AA) &  $\mu_{\rm Mn}$ ($\mu_{B}$) & $\mu_{\rm R}$ ($\mu_{B}$) \\
\hline
Tb &  4 (3.87) & 135 (10) & 20 (2) & 1.7 (3) & 3.5 (5) \\
Tb &  1 (1.04) & 200 (10) & 30 (3) & 2.5 (3) & 8.8 (8) \\
Y   & 4 (3.81) & 250 (10) &  13 (2) & 1.8 (2) & -- \\
Y   & 1 (1.38) & 190 (10) &  15 (3) & 3.0 (3) & -- \\
\end{tabular}
\end{center}
\caption{
\small{Transition temperature T$_{\rm SRO}$, correlation length  L$_c$ and
short range ordered magnetic moments at low temperature (8-11 K) deduced from neutron
diffraction data in the R(Mn$_{0.9}$Al$_{0.1}$)$_2$D$_y$ samples.  Error
bars are mentioned in units of the last digit.}}
\label{table1}
\end{table}

\begin{table}[!b]
\begin{center}
\begin{tabular}{ccc}
{\small Compound} & \multicolumn{2}{c}{\small  Effective magnetic moment ($\mu_{B}$)} \\
           & {\footnotesize (Magnetization)} & {\footnotesize 
           (Neutron scattering)}\\
\hline
Y(Mn$_{0.9}$Al$_{0.1}$)$_2$D$_{4}$ & 4.1 $\pm$ 0.2  & 4.0 $\pm$ 0.6 \\
Tb(Mn$_{0.9}$Al$_{0.1}$)$_2$D$_{4}$  & 10.1 $\pm$ 0.5 & 11.9 $\pm$ 1.8  \\
Tb(Mn$_{0.9}$Al$_{0.1}$)$_2$D$_{1}$ & 10.2 $\pm$ 0.5 & 11.6 $\pm$ 1.7 \\
\end{tabular}
\end{center}
\caption{
\small{Effective magnetic moments per chemical fomula in the
paramagnetic regime, deduced from magnetic and inelastic neutron scattering
measurements.}}
\label{table2}
\end{table}























\end{document}